\documentclass[prl,twocolumn,superscriptaddress]{revtex4}%

\usepackage[ansinew]{inputenc}
\usepackage{graphicx}
\usepackage{amsmath}
\usepackage{amsthm}
\usepackage{layout}
\usepackage{float}
\usepackage{subfigure}
\usepackage{amsfonts}
\usepackage{amssymb}
\usepackage{txfonts}

\begin{document}

\title{On-chip interaction-free measurements via the quantum Zeno effect\\}
\author{Xiao-song Ma}\affiliation{Department of Electrical Engineering, Yale University, New Haven, Connecticut 06511, USA}\affiliation{Institute for Quantum Optics and Quantum Information (IQOQI), Austrian Academy of Science}
\author{Xiang Guo}\affiliation{Department of Electrical Engineering, Yale University, New Haven, Connecticut 06511, USA}
\author{Carsten Schuck}\affiliation{Department of Electrical Engineering, Yale University, New Haven, Connecticut 06511, USA}
\author{King Y. Fong}\affiliation{Department of Electrical Engineering, Yale University, New Haven, Connecticut 06511, USA}
\author{Liang Jiang}\affiliation{Department of Applied Physics, Yale University, New Haven, Connecticut 06511, USA}
\author{Hong X. Tang}\affiliation{Department of Electrical Engineering, Yale University, New Haven, Connecticut 06511, USA}

\begin{abstract}
Although interference is a classical-wave phenomenon, the superposition principle, which underlies interference of individual particles, is at the heart of quantum physics. An interaction-free measurements (IFM) harnesses the wave-particle duality of single photons to sense the presence of an object via the modification of the interference pattern, which can be accomplished even if the photon and the object haven't interacted with each other. By using the quantum Zeno effect, the efficiency of an IFM can be made arbitrarily close to unity. Here we report an on-chip realization of the IFM based on silicon photonics. We exploit the inherent advantages of the lithographically written waveguides: excellent interferometric phase stability and mode matching, and obtain multipath interference with visibility above $98\%$. We achieved a normalized IFM efficiency up to $68.2\%$, which exceeds the $50\%$ limit of the original IFM proposal.
\end{abstract}


\maketitle

In classical physics, measurement processes require the interaction between the measurement device and the object to be measured. In quantum physics on the other hand, one can realize so-called interaction-free measurements (IFM), in which a measurement can be made without any physical interaction. The concept of IFMs was firstly considered by Dicke~\cite{Dick1981}. Elitzur and Vaidman (EV) extended this idea and proposed a gedanken experiment~\cite{Elitzur1993,Vaidman1994}. The goal of this gedanken experiment is to identify the presence of an ultra-sensitive bomb (i.e. any interaction with the bomb triggers an explosion) without causing it to explode. To achieve this goal, EV ingeniously proposed a quantum mechanical method by using the wave-particle duality of single photons. They proposed to incorporate the bomb into a Mach-Zehnder Interferometer (MZI) for achieving the IFM. The presence of a bomb modifies the optical interferograms of the MZI, even though photons and the bomb never interacted. By using this method, the ultra-sensitive bomb can be found without triggering it. The detail of their proposal is the following.

\begin{figure}[h]
\centering
\includegraphics[width=0.45\textwidth]{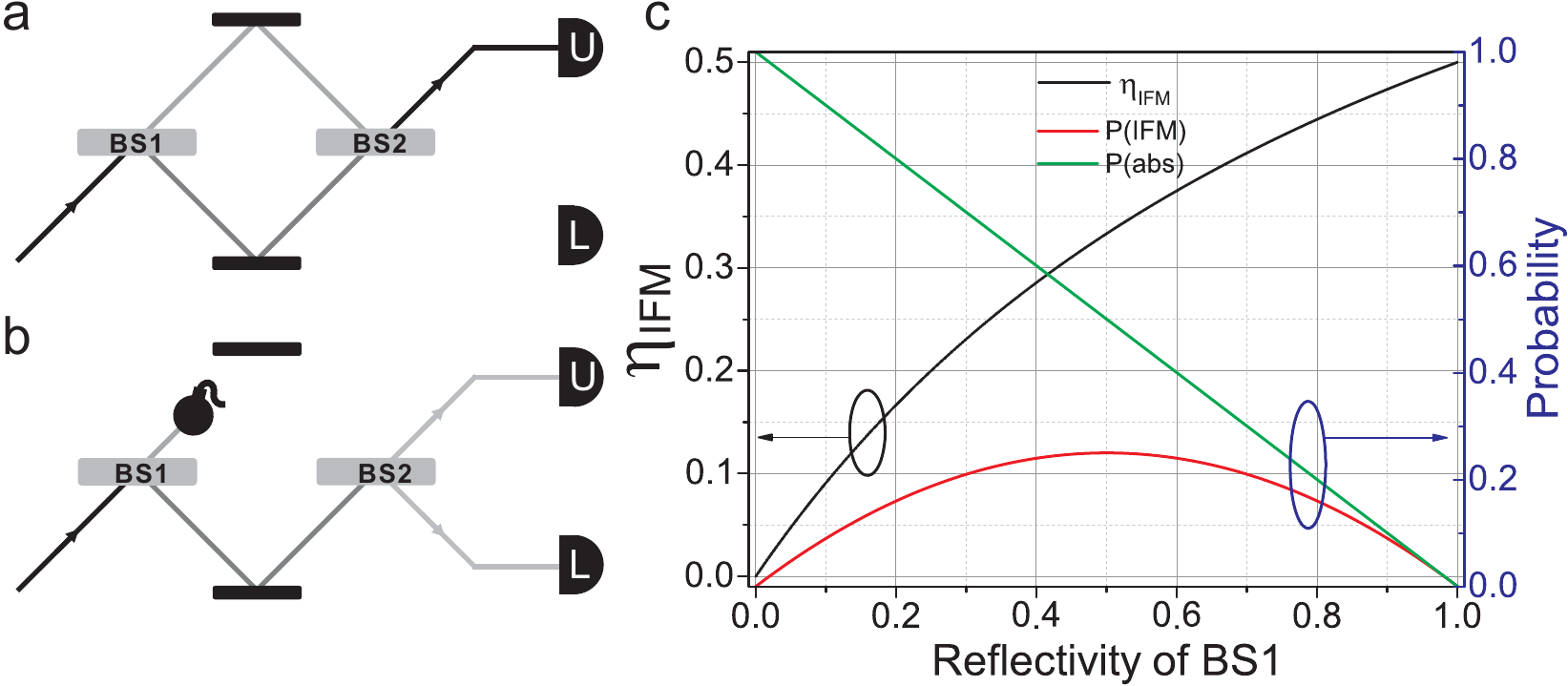}
\caption{The interaction-free measurement. \textbf{a}. An optical MZI is formed by two beam splitters (BS1 and BS2) and two mirrors. The MZI is aligned such that all the photons will go to Detector \textbf{U} and none to Detector \textbf{L}. \textbf{b}.  If the bomb is inserted, its presence will destroy the destructive interference at Detector \textbf{L}. Each detection event by \textbf{L} indicates the presence of the bomb and IFM succeeds. \textbf{c}. The efficiency of the successful IFM ($\eta_{IFM}$), the probabilities of successful IFM ($\textrm{P}(\textrm{IFM})$) and absorbtion of the photon by the bomb ($\textrm{P}(\textrm{abs})$) are shown in black, red and green curves, respectively.} \label{concept_EV}
\end{figure}

As shown in Fig.\ \ref{concept_EV}\textbf{a}, the relative phase between two arms of the MZI is adjusted to be zero such that any photon entering the MZI from the lower input will go to Detector \textbf{U} and none to Detector \textbf{L} due to constructive and destructive interference respectively. The probability of photon detection by Detector U, $\textrm{P}(\textrm{U})$, is 1 and that by Detector L, $\textrm{P}(\textrm{L})$, is 0.

If a bomb is inserted in the upper arm (Fig.\ \ref{concept_EV}\textbf{b}), the previous destructive interference at \textbf{L} is destroyed. The interaction between the photon and the bomb is \textit{not} required. The mere presence of the ultra-sensitive bomb destroys the coherence between the path states of the photon and inhibits the interference. Consequently, $\textrm{P}(\textrm{L})$ will not be zero any more. Any detection event by Detector L \textit{unambiguously} indicates the presence of the bomb. Moreover, a single photon is indivisible stemming from its particle property and cannot be split on a beam splitter~\cite{Grangier1986}. Therefore, every single photon detected by \textbf{L} must have propagated through the lower arm of the MZI and hence hasn't interacted with the bomb. Every single detection event by \textbf{L} is a successful IFM. The probability of successful IFM, $\textrm{P}(\textrm{IFM})$, equals $\textrm{P}(\textrm{L})$. The detection events at detector \textbf{U} are inconclusive as they don't tell us whether the bomb is present or not. The input photon can also be absorbed by the bomb, trigging an explosion, with a probability of $\textrm{P}(\textrm{abs})$. To quantify the performance of an IFM, an efficiency parameter, $\eta_{IFM}$, is introduced as the fraction of conclusive measurements which are interaction free:
\begin{equation}\label{eta}
\eta_{IFM}=\frac{\textrm{P}(\textrm{L})}{\textrm{P}(\textrm{abs})+\textrm{P}(\textrm{L})}=\frac{R_{BS1}R_{BS2}}{T_{BS1}+R_{BS1}R_{BS2}},
\end{equation}
where $R_{BS1}/R_{BS2}$ and $T_{BS1}/T_{BS2}$ are the reflectivity and transmissivity of BS1/BS2. If we use two balanced beam splitters, i.e. $R_{BS1}=R_{BS2}=0.5$, $\eta_{IFM}$ will be $1/3$. EV further showed by changing the beam splitter's reflectivity one could increase the efficiency of the IFM to $1/2$ (shown in Fig.\ \ref{concept_EV}\textbf{c}).

It is essential to use a pair of complementary beam splitters, i.e. $T_{BS1}=R_{BS2}$ and hence $R_{BS1}=T_{BS2}$ (Fig.\ \ref{concept_EV}\textbf{b}). Given zero relative phase of the MZI, the employment of a pair of complementary beam splitters ensures perfect destructive interference at \textbf{L} when the bomb is absent. This configuration guarantees that when the bomb is present, any detection event in \textbf{L} unambiguously indicates a successful IFM and $\eta_{IFM}=R_{BS1}/(1+R_{BS1})$.

Note that the visibility between \textbf{U} and \textbf{L} without bomb is $\frac{(\textrm{P}(\textrm{U})-\textrm{P}(\textrm{L}))}{(\textrm{P}(\textrm{U})+\textrm{P}(\textrm{L}))}$ and gives the confidence level of the success of IFM given a detection event of \textbf{L} when the bomb is present. There have been several experimental realizations of EV's IFMs in different physical systems~\cite{Kwiat1995a,Voorthuysen1996,Hafner1997}. Also, ``interaction-free" imaging was reported \cite{White1998}. Although EV's proposal is elegant and makes impossible in classical physics possible in quantum mechanics, it has two limitations (shown in Fig.\ \ref{concept_EV}\textbf{c}): (1) The efficiency of an IFM, $\eta_{IFM}$, has an upper bound of $1/2$. (2) When R=1, $\eta_{IFM}$ is at its maximum $1/2$. But the probability of IFM, $\textrm{P}(\textrm{IFM})$ is arbitrarily close to 0 and the inconclusive measurement $\textrm{P}(\textrm{U})$ is close to unity. For the practical IFMs, it is of crucial importance to increase $\eta_{IFM}$ and $\textrm{P}(\textrm{IFM})$ in the same time.

% To some extent, it is similar to an unbiased guess.
%In the connected Mach-Zehnder interferometers (cMZI) with N beam splitters, each beam splitter's reflectivity equals $cos^{2}(\frac{\pi}{2N})$.
%Each detection event by \textbf{L} signals an IFM.

\begin{figure}[h!]
\centering
\includegraphics[width=0.5\textwidth]{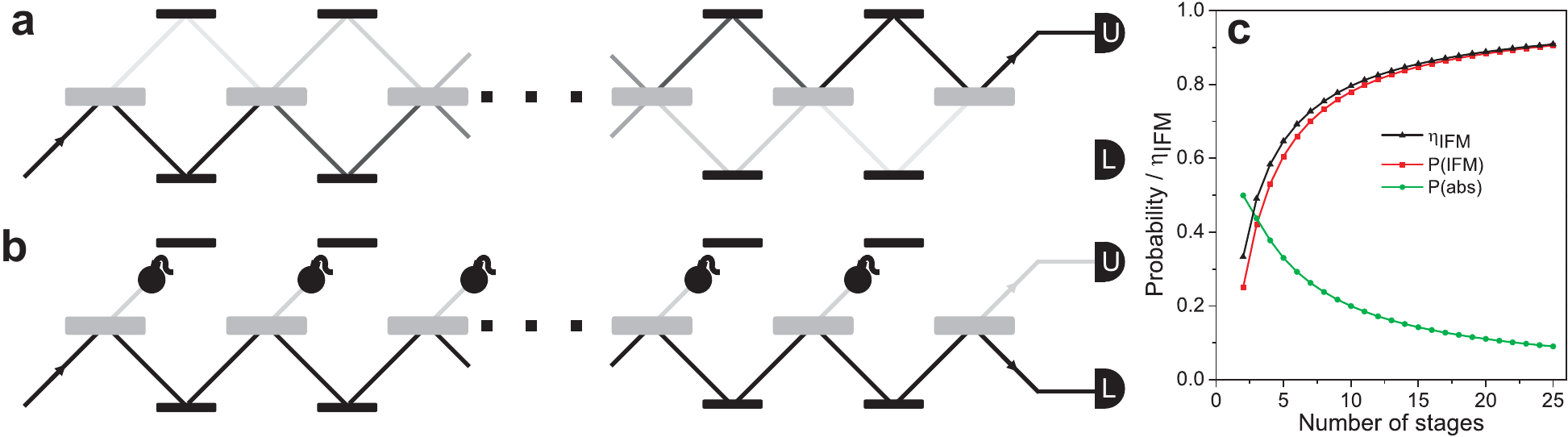}
\caption{Quantum Zeno effect enabled interaction-free measurement. \textbf{a}. A photon entering the cMZI from the lower input will coherently and gradually evolve to the upper half of the cMZI, and is then detected by \textbf{U}. \textbf{b}. If the bombs are inserted in the upper arm of the cMZI, the multi-stage interference is destroyed and \textbf{L} will detect a large portion of the input photons due to the quantum Zeno effect. \textbf{c}. By increasing the number of the beam splitters, both $\eta_{IFM}$ (black triangles) and $\textrm{P}(\textrm{IFM})$ (red squares) can be made arbitrarily close to 1, and $\textrm{P}(\textrm{abs})$ (green disks) is reduced to close to 0.} \label{concept_QZIFM}
\end{figure}

In order to enhance the efficiency of an IFM, Kwiat \textit{et al}. combined the discrete form of the quantum Zeno effect~\cite{Misra1977,Peres1980} with IFMs, where one coherently and repeatedly probes a region that might contain the bombs~\cite{Kwiat1995a,Kwiat1999}. This quantum Zeno effect enabled IFM (QZIFM) in principle allows both $\eta_{IFM}$ and $\textrm{P}(\textrm{IFM})$ to approach unity and hence allows the detection of ultra sensitive bombs with an arbitrarily small chance of triggering them (absorbing a photon). The scenario without bombs is depicted in Fig.\ \ref{concept_QZIFM}(\textbf{a}). A photon enters the connected Mach-Zehnder interferometers (cMZI) and its path state will \textit{gradually} and \textit{coherently} evolve from the lower half to the upper half of the cMZI. If all the beam splitters' reflectivities fulfills $R=cos(\pi/(2N))^2$ (where $N$ is the number of beam splitters), the photon will exit via the upper port of the final beam splitter with certainty after all $N$ stages, i.e. $\textrm{P}(\textrm{U})=1$. The photon has zero probability to exit from the lower port, i.e. $\textrm{P}(\textrm{L})=0$.

As shown in Fig.\ \ref{concept_QZIFM}(\textbf{b}), if bombs are inserted into the upper part of the cMZI, the photon's coherent evolution is inhabited and it will propagate through the lower part of the cMZI with a probability $\textrm{P}(\textrm{L})=[cos^{2}(\frac{\pi}{2N})]^{N}$ of being detected by \textbf{L}, whereas this probability was 0 when there were no bombs. The probability of the photon being detected by the upper detector \textbf{U} is: $\textrm{P}(\textrm{U})=[cos^{2}(\frac{\pi}{2N})]^{N-1}sin^{2}(\frac{\pi}{2N})$. Assuming the cMZI is lossless, the absorbtion probability is $\textrm{P}(\textrm{abs})=1-\textrm{P}(\textrm{L})-\textrm{P}(\textrm{U})$. As shown in Fig.\ \ref{concept_QZIFM}(\textbf{c}), one can see both $\eta_{IFM}$ and $\textrm{P}(\textrm{IFM})$ increase as $N$ increases and can be arbitrarily close to one in the limit of large $N$. These are the unique advantages of a QZIFM as compared to the original IFM.

Since the QZIFM was proposed~\cite{Kwiat1995a}, there have been several endeavours in realizing it, including a broadband, discrete method~\cite{Kwiat1999}, resonant, continuous methods~\cite{Paul1996,Tsegaye1998} and both~\cite{Streed2006}. The quantum Zeno effect is also essential in certain quantum computation schemes~\cite{Franson2004}, counterfactual quantum computation~\cite{Mitchison2001,Hosten2006},  quantum state protection~\cite{Paz-Silva2012} and all-optical switching~\cite{Wen2012,McCusker2013}. However, the challenges of the previous demonstration with light~\cite{Kwiat1999} are the noise caused by interferometric (sub-wavelength) instability, despite active stabilization, and imperfect optical elements.

% INTEGRATED QUANTUM PHOTONICS

Integrated quantum photonics is a promising approach to realize quantum information processing, as it offers interferometric-stable, miniature and scalable solutions due to its monolithic implementation~\cite{Metcalf2013,Thompson2011,Brien2009}. The silicon-on-insulator (SOI) platform is particularly attractive because: (1) it provides good mode confinement due to high refractive index contrast; (2) well-established fabrication techniques allow to implement complex quantum circuits; (3) it is compatible with superconducting material which enabled the realization of waveguide-integrated single-photon detectors~\cite{Pernice2012,Schuck2013a,Schuck2013b,Schuck2013c}; (4) on-chip quantum interference between single photons~\cite{Xu2013} and photon-pair sources~\cite{Silverstone2014} have been recently realized.

Here we demonstrate discrete quantum Zeno effect enabled IFMs up to 20 stages on the SOI platform. We employ direct write lithography to realize the circuit conceptually shown in~\cite{Kwiat1995a} and Fig.\ \ref{concept_QZIFM} for realizing a discrete QZIFM without the need of self-stabilized interferometers or active phase stabilization, which greatly enhances the practicality of the implementation of IFM. The detailed information on the design and fabrication of our devices can be found in ref. ~\cite{SM2014}.

\begin{figure}[t]
\centerline{\includegraphics[width=0.5\textwidth]{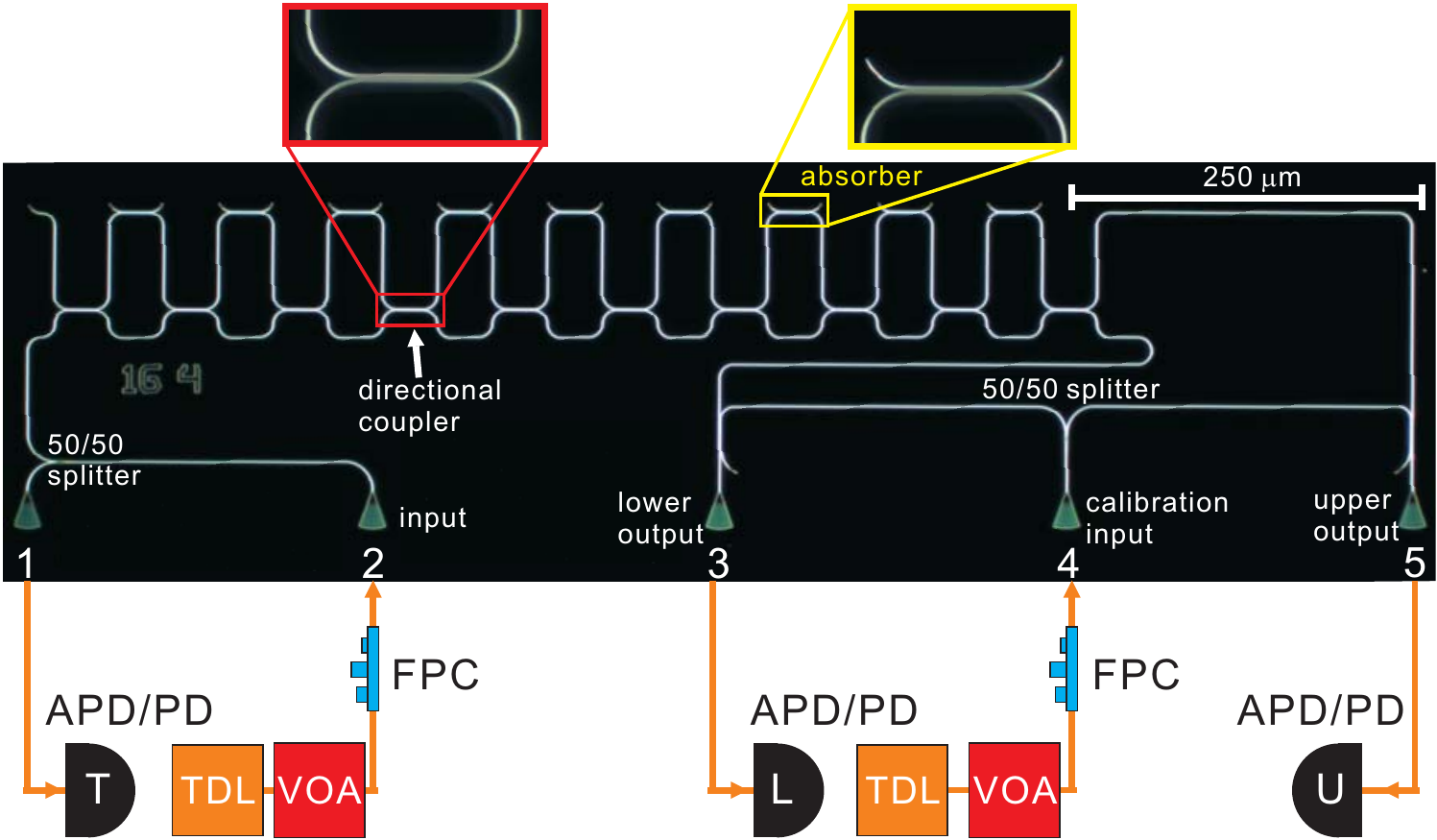}}
\caption{The optical micrograph and schematic setup of a device with 10-stage QZIFM circuitry. There are five grating couplers (GC, labeled in numbers). GC 2 is used as the input. GC 1 couples out half of the total input power, which is measured by Detector T. GCs 3 and 5 couple out the lower and upper outputs of the MZI circuit, which are then measured by Detectors L and U respectively. A tunable diode laser (TDL) is used as the light source and a variable optical attenuator (VOA) is used to attenuate the laser to single-photon level. A fiber polarization controller (FPC) is used to rotate the polarization of the input light to be TE. A set of same TDL, VOA and FPC provides light source to GC 4 for measuring the output coupling differences between GCs 3 and 5. Scale bar, 250 $\mu$m.} \label{layout}
\end{figure}

Our experimental setup is depicted in Fig.\ \ref{layout}. To characterize the device, we use a telecom tunable diode laser (TDL) as the light source. In order to launch light into and collect the output from the QZIFM circuitry under test, a single-mode fibre array with a pitch of 250 $\mu$m is used. On the chip, we use grating couplers~\cite{Taillaert2004, Li2008} as optical input/output ports.

First we implemented the two-stage IFM depicted in Fig.\ \ref{concept_EV}. To characterize the performance of the device, we set the path-length of the upper arm to be $100~\mu m$ longer than the lower one and measure the transmission spectra with a tunable diode laser and linear photodetectors~\cite{SM2014}. We supply either pseudo single photons for IFM demonstrations or laser light for device characterizations to port 2. Then the signals are split with a 50/50 splitter. Half of the light is directed to port 1, where it is coupled out into an optical fiber and measured by Det(T). The other half is sent to the QZIFM device. The lower and the upper outputs from the QZIFM device are directed to port 3 and 5 and are then guided with optical fibers to Det(L) and Det(U), respectively.

\begin{figure}[t]
\centerline{\includegraphics[width=0.45\textwidth]{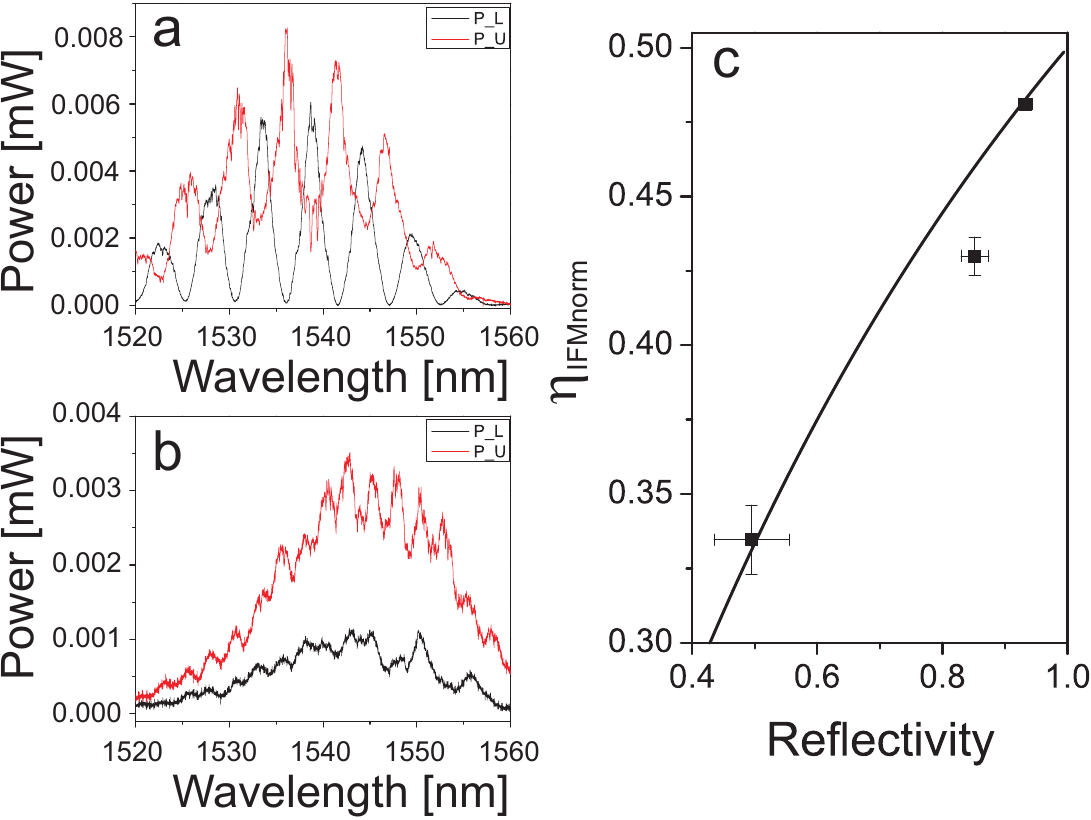}}
\caption{Experimental data of IFM. \textbf{a} The transmission spectra of the lower and upper outputs of a device without absorber (corresponding to Fig.\ \ref{concept_EV}\textbf{a}). \textbf{b} The transmission spectra with a full absorber being present in the upper arm of the interferometer. \textbf{c}. The result of the 2-stage IFM as a function of reflectivity of the first directional coupler. The black curve is the theoretical prediction as in Fig.\ \ref{concept_EV}\textbf{c}.} \label{data_EV}
\end{figure}

We introduced another tapered waveguide as the absorber/``bomb", which lies close to the upper arm of the interferometer. When the absorber waveguide is positioned 10 $\mu m$ away from the upper arm of the interferometer, there is negligible coupling between them. Hence, this situation represents the case without absorber. In this case, we experimentally obtain high-contrast transmission spectra at the upper and the lower outputs with above $99.8\%$ interference visibility at the wavelength that corresponds to the phase of being the multiples of $2\pi$ (at wavelength of 1541.49 nm). The results of a device with $0.852\pm0.022$ reflectivity at the first directional coupler are shown in Fig.\ \ref{data_EV}\textbf{a}. The reflectivities' uncertainties stem from the uncertainties in the waveguide width control ($\pm10 nm$) during fabrication process. The upper output's minimum doesn't go to zero because the directional couplers' reflectivities are not 0.5.

To demonstrate an IFM, the gap between absorber and upper arm of MZI is set such that all the light in the upper arm couples to the absorber waveguide and hence this absorber is a full absorber. In our case, the gap of a full absorber is about 190 nm. The presence of this full absorber destroys the high-contrast interference. In Fig.\ \ref{data_EV}\textbf{b}, we show the transmission spectra where the full absorber is present. Note that other than the gap size between absorber and MZI, this device has the identical nominal design as the one shown in Fig.\ \ref{data_EV}\textbf{a}. It is clearly visible that the interference patterns disappeared. In order to demonstrate an IFM, the laser was attenuated such that the average photon number is about 0.1 per gate and the output photons are measured with single-photon detectors made by InGaAs avalanche photodiodes with 100-kHz gate.

We note that in our experiment true single photons and photons from laser being attenuated to single-photon level behave similarly because only single-photon interference and linear optics are concerned. It is noteworthy that by using photons from laser being attenuated to single-photon level, there will be a very small chance (less than 0.05 in our case) that two photons exist simultaneously per gate. These two photons might cause the explosion of the bomb as well as the detection events of lower detector with a small chance. This is similar to use weak coherent laser pulses to implement quantum cryptography~\cite{Gisin2002}, in which the communication security is sensitive photon number splitting attack.

The IFM's efficiencies are derived from the photon counts and shown in Fig.\ \ref{data_EV} \textbf{c}. Based on the high-visibility interference, we believe the first directional coupler (DC) and the second DC are complementary to each other, i. e. the reflectivity of the first DC ($R_{DC1}$) equals to the transmissivity of the second DC ($T_{DC2}$). We derive the normalized efficiency of the IFM from:
$\eta_{\textrm{IFMnorm}}=\frac{T_{DC2}}{1+T_{DC2}}=\frac{1}{2+\frac{C_L}{C_U}\frac{a_U}{a_L}}$. $C_L$, $C_U$, $a_U$, and $a_L$ are the single counts and coupling (as well as detecting) efficiencies from the lower output and upper output from the device, respectively. By using the calibration input GC 4 and 50/50 splitter, we obtained $\frac{a_L}{a_U}$ and its uncertainties. They are measured at wavelength of 1541.49 nm. Note that in deriving the normalized efficiency of the IFM, we have factored out the coupling and detecting efficiencies. To improve the efficiency of the IFM, it is necessary to include high-efficiency on-chip single-photon detectors~\cite{Pernice2012}. The error bars include both systematic (due to fabrication inhomogeneities) and statistical errors assuming Poissonian statistics.

Next we employ the quantum Zeno effect to enhance the efficiency of IFM and fabricate devices with multiple connected interferometers as schematically depicted in Fig.\ \ref{concept_QZIFM}. A ten-stage QZIFM device is shown in Fig.\ \ref{layout}. As mentioned previously, the reflectivity of each directional coupler should be set to be $R=cos(\pi/(2N))^2$. Only when this condition is fulfilled, the path state of the photon will coherently evolve from the lower path to the upper path after $N$ stages, as shown in Fig.\ \ref{concept_QZIFM}\textbf{a}. For $N=5$, $10$ and $20$, the reflectivities of each directional coupler are about 0.904, 0.975 and 0.994, respectively.

\begin{figure}[t]
\centerline{\includegraphics[width=0.45\textwidth]{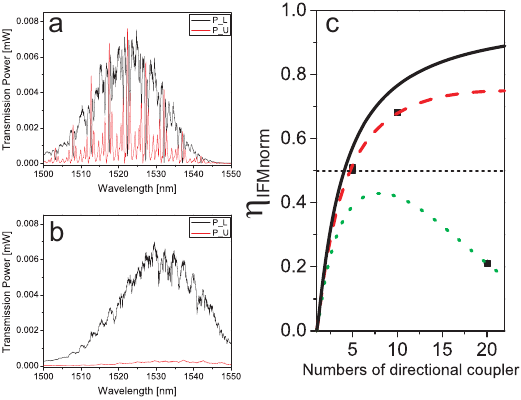}}
\caption{Experimental data of the QZIFM. \textbf{a} shows the transmission spectra of the lower and upper outputs of a 10-stage device without absorber. \textbf{b}. The transmission spectra with 9 full absorbers being present in the upper arm of the interferometer. \textbf{c}. The results of the 5-, 10- and 20-stage QZIFM. The solid black curve is the theoretical prediction of a lossless device. The red dashed and green dotted curves are the predictions with $7.4\%$ and $21.2\%$ loss per stage. Data of $N=5$ (measured at the wavelength of 1539.75 nm) and $10$ (measured at the wavelength of 1527.25 nm) are in good agreement with the red dashed curve. Data of $N=20$ (measured at the wavelength of 1538.645 nm) show higher loss and is in agreement with the green curve~\cite{SM2014}.} \label{data_QZ}
\end{figure}

The measured transmission spectra are shown in Fig.\ \ref{data_QZ}\textbf{a}. Excellent interference with more than $98\%$ visibility has been obtained. Our lithographically defined circuitries provided perfect spatial mode matching and stable phase, which are difficult to achieve with traditional bulk optics especially for the multistage interferometers. This shows that our system is a very good platform for the development of waveguide quantum optic circuits. Note there are several side peaks with low amplitudes between the main ones with high amplitudes. This is because that as the number of directional couplers increases; multi-path interference occurred and hence complicated interference patterns show up.

In the case of QZIFM, we positioned the absorber about $190$ nm away from the upper arm of the interferometer, such that all the light is coupled to the absorber from the upper arm. This corresponds to the scenario depicted in Fig.\ \ref{concept_QZIFM}\textbf{b}. In this case, we expect to obtain high output in the lower arm and low output in the upper arm. We present the laser characterization of a QZIFM device with absorbers in Fig.\ \ref{data_QZ}\textbf{b}. We clearly observe that the high-contrast interference disappears. The small modulations visible in Fig.\ \ref{data_QZ}\textbf{c} originate from the Fabry-P\'{e}rot interferometer formed by input and output grating couplers as confirmed with independently tested calibration devices.

Fig.\ \ref{data_QZ}\textbf{c} shows the normalized efficiencies of QZIFM devices with different numbers of directional couplers. We obtained these results by using a strongly attenuated laser as the light source and single-photon detectors (the same as Fig.\ \ref{data_EV} \textbf{c}). Here we derive the normalized efficiencies of IFM via $\eta_{IFMnorm}=\frac{C_L}{C_T-C_U \cdot \frac{a_L}{a_U}}$, where $C_T$, $C_L$ and $C_U$ are the single-photon counts from GC 1, 3 and 5. Note that we assume the coupling and detecting efficiencies are the same for T and L outputs because we use the same fiber and detector to measure $C_T$ and $C_L$. We obtained $\eta_{\textrm{IFM}}$ of $0.506\pm0.014$, $0.682\pm0.008$ and $0.212\pm0.002$ for $N=5$, $10$ and $20$, respectively. The solid black curve is the theoretical prediction of a lossless device. The red dashed and green dotted curves are the predictions with $7.4\%$ and $21.2\%$ loss per stage~\cite{Kwiat1998,Kwiat1999}. Data of $N=5$ and $10$ are in good agreement with the red dashed curve. We note that for larger $N$ , the device becomes so long that we have to define the lithographic pattern over multiple electron beam write fields. Hence, we attribute the extra loss to the stitching error between waveguides in different write fields~\cite{SM2014} and the build-up of the mode-conversion loss in the coupling regions~\cite{Xia2006}.

In conclusion, we report the realizations of interaction-free measurement via quantum zeno effect on a silicon photonic chip. The future direction would be to further enhance the efficiency of IFM with lower-loss circuitry~\cite{Vlasov2004} and on-chip single-photon detectors~\cite{Pernice2012}. Additionally, by using micro-ring resonators, it is possible to further enhance the efficiency of IFM as well as to interrogate the presence of a single absorber via multiple passages, which could be useful for certain practical implementations. Our realizations of the interaction-free measurement via quantum zeno effect could be useful in spectroscopic studying photosensitive materials.

\begin{acknowledgements}
We are grateful to Xufeng Zhang, Xiong Chi and Changling Zou for discussions and Alexander V. Sergienko for lending us the single-photon detectors. X.S.M. thanks J. Kofler for discussions. X.S.M is supported by a Marie Curie International Outgoing Fellowship within the $7^{th}$ European Community Framework Programme. C.S. gratefully acknowledges financial support from the Deutsche Forschungsgemeinschaft (DFG-Forschungsstipendium). L.J. acknowledges support from the Alfred P Sloan Foundation, the Packard Foundation, the AFOSR-MURI, the and DARPA Quiness program. H.X.T. acknowledges support from a Packard Fellowship in Science and Engineering and a CAREER award from the National Science Foundation. We thank Dr. Michael Rooks and Michael Power for their assistance in device fabrication.
\end{acknowledgements}

\bibliographystyle{apsrev}

\section{Supplemental Material}
In the Supplemental Material, we first provide a detailed description on the design of the optical waveguides the directional couplers. Then we present the characterizations of the on-chip interferometers. Finally, we show the data on the quantum Zeno effect enabled interaction-free measurement device with 20 directional couplers.

\section{Design and fabrication of the devices}
% the reflectivity vs gap.

Waveguides are fabricated on a SOI wafer, which has a 220 nm thick layer of silicon on top of a 3 $\mu$m thick buried oxide layer that prevents the optical modes from leaking to the substrate. The width of the waveguides is chosen to be 400 nm to ensure: (1) single-mode propagation of the transverse-electric (TE) polarized mode; (2) a short coupling length of about 20 $\mu$m for various directional couplers, thus achieving small device footprint; (3) low transmission loss. The gap for evanescent coupling between two waveguides building up a directional coupler is chosen according to the desired beam splitter reflectivity. The bending radius in our device is chosen to be 10 $\mu$m in order to guarantee low bending loss~\cite{Vlasov2004}. Waveguides, directional couplers and grating couplers are defined by electron beam lithography in hydrogen silsesquioxane resist and subsequently etched in an inductively-coupled chlorine plasma reactive ion etch.

\begin{figure}[h!]
\centerline{\includegraphics[width=0.48\textwidth]{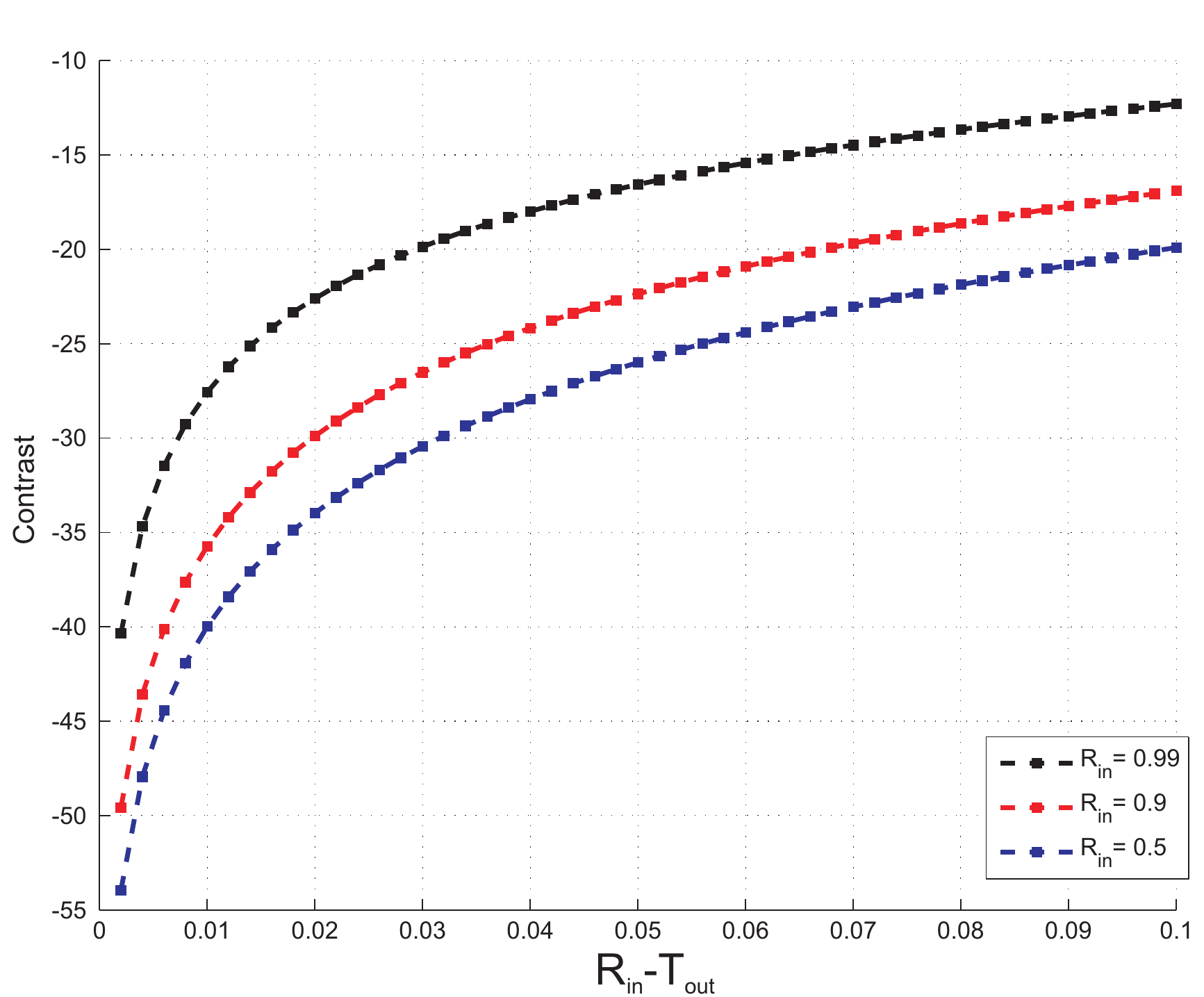}}
\caption{In the interaction-free measurement with a two-stage Mach-Zehnder interferometer, it is crucial to ensure that the input and output directional couplers are complementary to each other in order to obtain high-contrast interference. This means the transmissivity of output directional coupler ($T_{out}$) should equal the reflectivity of the input directional coupler ($R_in$), i. e. $R_{in}=T_{out}$ or equivalently $R_{out}=T_{in}$. The interference contrast is defined as the ratio between the minimum intensity (count rate) of the lower output , \textbf{L} (see Fig. 1 of the main text), and the maximum of the upper output, \textbf{U}, for a given phase. We vary $T_{out}$ for $R_{in}=0.99$ (black), $0.9$ (red) and $0.5$ (blue) and obtain that the interference contrast decreases as $R_in-T_out$ increases.} \label{mismatch}
\end{figure}

In our experiment, it is crucial to obtain the desired reflectivities/transmissivities of the directional couplers (DC) accurately, because the visibility of the transmission spectra is sensitive to these parameters. In the interaction-free measurement (IFM) with a regular two-stage Mach-Zehnder interferometer, we have to ensure the input and output directional couplers have complementary reflectivities, which means that the transmissivity of the output directional coupler ($T_{out}$) should equal the reflectivity of the input directional coupler ($R_{in}$). Any mismatch between input reflectivity and output transmissivity will reduce the interference contrast. In order to show this effect, in Fig.\ \ref{mismatch}, we vary $T_{out}$ and plot the interference contrast versus the mismatch ($R_{in}-T_{out}$) for three different $R_{in}$: 0.99, 0.9 and 0.5. We consider output transmissivities corresponding to a mismatch in the range $R_{in}-T_{out}=0.1$, which is reasonable based on the uncertainties in fabrication process. Note that a mismatch of 0 corresponds to the perfectly matched case.

\begin{figure}[ht!]
\centerline{\includegraphics[width=0.5\textwidth]{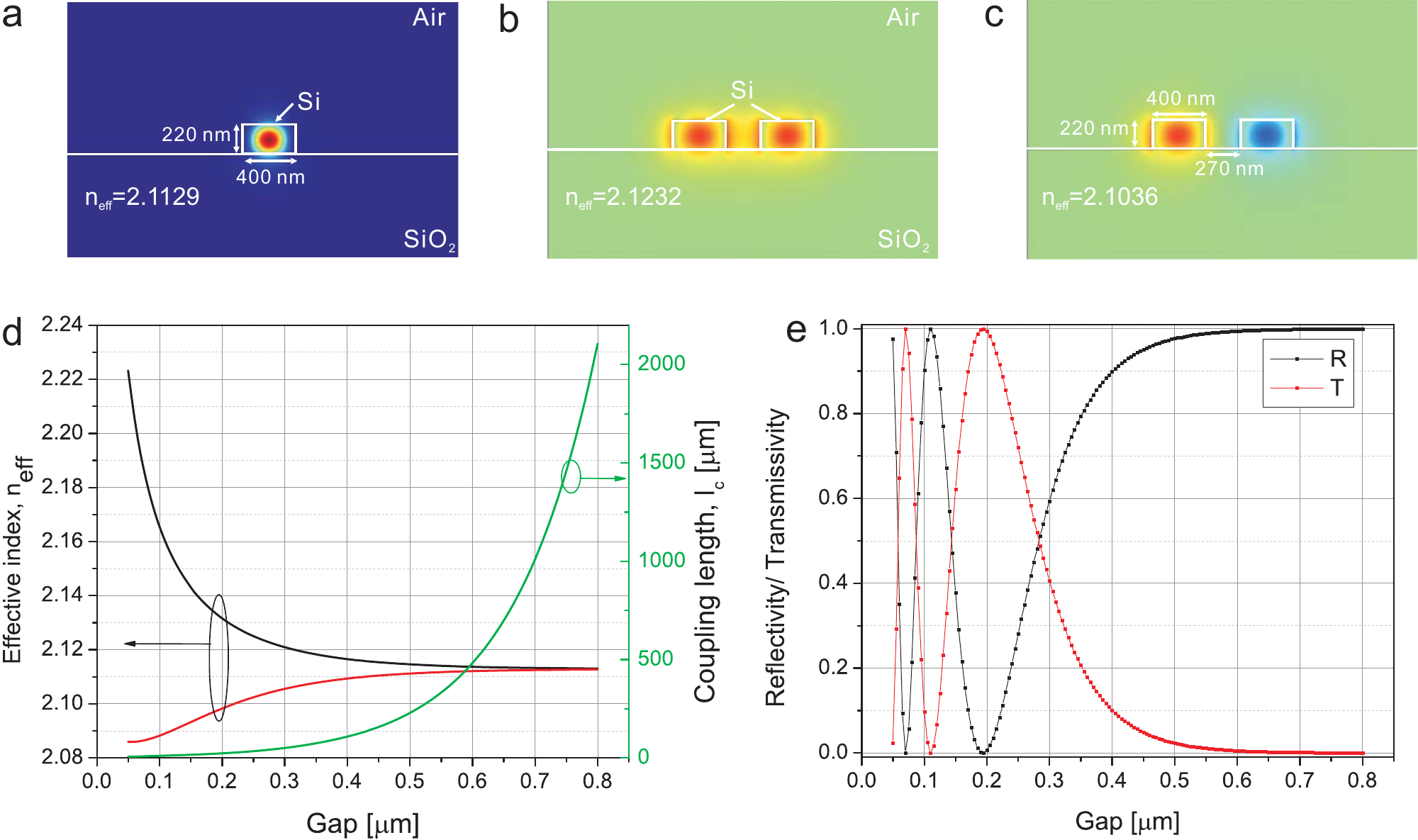}}
\caption{Design of directional couplers. \textbf{a}. The normalized power distribution of the propagating transverse electric (TE) mode confined by a single Si waveguide with 400~nm/220~nm in width/thickness. The effective refractive index ($n_{eff}$) of this mode is about 2.1129. In the coupling region, two waveguides are brought close to each other and mode hybridization occurs due to the evanescent coupling. Therefore, new compound modes become the new eigenmodes. \textbf{b} and \textbf{c}. The simulated TE-like electric field component, $E_x$ of the symmetric and the antisymmetric compound modes. In this simulation, we choose the gap between the two waveguides to be 270 nm. For the symmetric/antisymmetric mode, the electrical fields distributed in the two silicon waveguides are in/out of phase, i.e. the relative phase is 0/$\pi$. The effective refractive index ($n_{eff}$) of the symmetric and antisymmetric modes are about 2.1232/2.1036. The power and electrical field distributions in \textbf{a}-\textbf{c} are shown in linear color scale. \textbf{d}. The effective refractive indexes of these two compound modes as functions of the gap between the two waveguides. The black curve is the effective index of the symmetric mode ($n_{s}$) and the red is that of the anti-symmetric mode ($n_{a}$). See text for details on how the coupling length of these two waveguides, $l_c$, can be derived. As we increase the gap, the difference of the effective refractive indexes of these two modes becomes smaller and hence the coupling length becomes larger. \textbf{e}. The reflectivity and transmissivity of a directional coupler composed by the above mentioned waveguides for a design length of 20 $\mu$m. Note that the incoming and outgoing bend regions visible in Fig 3 of the main text.} \label{design}
\end{figure}

In the case of a quantum-Zeno enabled interaction-free measurement (QZIFM) with $N$ directional couplers, the crucial point is to realize directional couplers with the reflectivity $R=cos(\pi/(2N))^2$. Only if this condition is fulfilled, Fig.\ 2(\textbf{a}) in the main text will be realized (absorbers are absent).

In the presence of absorber waveguides, it is furthermore crucial to make sure that the directional couplers formed by the absorber and the upper arm of the MZI has the reflectivity of 0. Any over- or under-coupling between the absorber and the upper arm of the interferometer will lead to unwanted interference and lower the confidence level of QZIFM (as explained in the main text).

In order to have a guidance in designing our circuitry, we used a waveguide mode solver (COMSOL) to calculate the effective refractive indexes of the optical modes confined by the waveguide structures. First, we simulate the fundamental transverse electric (TE) mode confined by a single Si waveguide with 400~nm/220~nm in width/thickness, shown in  Fig.\ \ref{design}\textbf{a}, and calculate the effective index and group index of this mode. In the directional coupler region, the two coupling waveguides are close to each other and mode hybridization occurs due to evanescent couplings~\cite{Little1995}. Therefore, new compound modes become the new eigenmodes of this coupled waveguides. In Fig.\ \ref{design}\textbf{b} and \textbf{c}, we plot the simulated distributions of the TE-like electric field component, $E_x$ of the symmetric and the antisymmetric compound modes. In this simulation, we choose the gap between the two waveguides to be 270 nm. For the symmetric/antisymmetric mode, the electrical fields distributed in the two silicon waveguides are in/out of phase, i.e. the relative phase is 0/$\pi$. In Fig.\ \ref{design}\textbf{d}, we plot the effective indexes of the symmetric ($n_{s}$) and anti-symmetric ($n_{a}$) modes as a function of the gaps between the waveguides. Based on these effective indexes, we can derive the coupling length of the two straight waveguides via $l_{c}=\lambda/2(|n_{s}-n_{a}|)$.  $l_{c}$ is the length over which the total amount of power is transferred from one waveguide to another.

The grating couplers' design is similar to that in ref~\cite{Li2008}. Here in order to remove unwanted oscillations from the Fabry-P\'{e}rot interferometer formed by the input and output grating couplers, we use an apodized design at the end of each grating couplers.

For device fabrication, we chose a fixed design length of the directional coupler ($l=20\mu m$) and varied the gap between the waveguides. This allowed us to change the effective refractive indexes of the symmetric and antisymmetric modes and hence vary $l_{c}$. The incoming and outgoing bend regions increase the coupling length about 2 $\mu$m in our case. The transmissivity and reflectivity of the directional couplers are $T=[sin(\frac{\pi l}{2 l_c})]^{2}$ and $R=[cos(\frac{\pi l}{2 l_c})]^{2}$, respectively, as shown in Fig.\ \ref{design}\textbf{e}. Note that in this simulation, we assume the side walls of the waveguides are vertical, which slightly deviates from out fabricated devices due to inhomogeneities in dry etching.

\section{Characterization of the on-chip interferometers}

% the connection between the phase and the wavelength

%We initially fabricated devices with identical arm length for the upper and lower arms. When we measure a device without absorbers and found the contrast is low, we have to distinguish whether this is due to the inaccurate reflectivities of the directional couplers or the phase of the interferometer is not multiples of $2\pi$. In order to separate these two possibilities,

In this section we will discuss the procedure for characterizing the optical performance of our devices. In out MZI devices, we included extra waveguide sections of length $\Delta L=100$ $\mu m$ in the upper arms, which act as "highly dispersive" elements in the interferometer~\cite{Vlasov2005,Dulkeith2006,Li2008}. This allows us to tune the phase by scanning the wavelength of the input and hence evaluate the performance of the devices by measuring the transmission spectra of the device. In the MZI case, the high-visibility interference in the transmission spectra signals that the first and the second directional coupler are complementary to each other, i. e. the reflectivity of the first DC equals to the transmissivity of the second DC.

To illustrate this, we consider a simple case of a two-stage interferometer, i. e. a Mach-Zehnder interferometer. The phase difference between the two arms, $\Delta\phi$, is given by:
\begin{equation}\label{phase_diffrence}
\Delta\phi=2\pi \Delta L\cdot (\frac{n_{eff}}{\lambda}),
\end{equation}
where $n_{eff}$ and $\lambda$ are the effective refractive index of the propagating mode in a single waveguide and the free-space wavelength. Since we are interested in the phase shift induced by the wavelength change, we can derive the differential phase shift per unit wavelength ($\frac{d(\Delta\phi)}{d\lambda}$):
\begin{equation}\label{diffrential_phase}
\frac{d(\Delta\phi)}{d\lambda}=2\pi \Delta L\cdot [\frac{1}{\lambda} \cdot (\frac{d n_{eff}}{d \lambda})-\frac{n_{eff}}{\lambda^2}]=-\frac{2\pi \Delta L n_g}{\lambda^2},
\end{equation}
where $n_g$ is the group index and can be calculated either from experimental data or from simulations. In our case, the group index at 1550 nm is about 4.7 from simulation. To calculate the free spectral range (FSR), i.e. the period of the interference pattern, we measure the wavelength difference between the nearest-neighbouring interference dips, $\textrm{FSR}=\lambda_2-\lambda_1$, with $\lambda_1 < \lambda_2$. The phase difference between these two wavelength is $2\pi$ and then we have:
\begin{equation}\label{FSR}
\frac{d(\Delta\phi)}{d\lambda}|_{\lambda=\lambda_2} \cdot \lambda_2-\frac{d(\Delta\phi)}{d\lambda}|_{\lambda=\lambda_1} \cdot \lambda_1=2\pi.
\end{equation}
Then by using Eq.\ \ref{diffrential_phase}, we arrive at:
\begin{equation}\label{FSR2}
(\frac{1}{\lambda_1}-\frac{1}{\lambda_2})=\frac{1}{n_g \Delta L}.
\end{equation}
Because $\lambda_1$ and $\lambda_2$ are close to each other, we here assumed that the group index of $\lambda_1$ and $\lambda_2$ are the same and equal to $n_g$.

% Data with the wrong reflectivity

\section{Various devices}
\begin{figure}[h!]
\centerline{\includegraphics[width=0.5\textwidth]{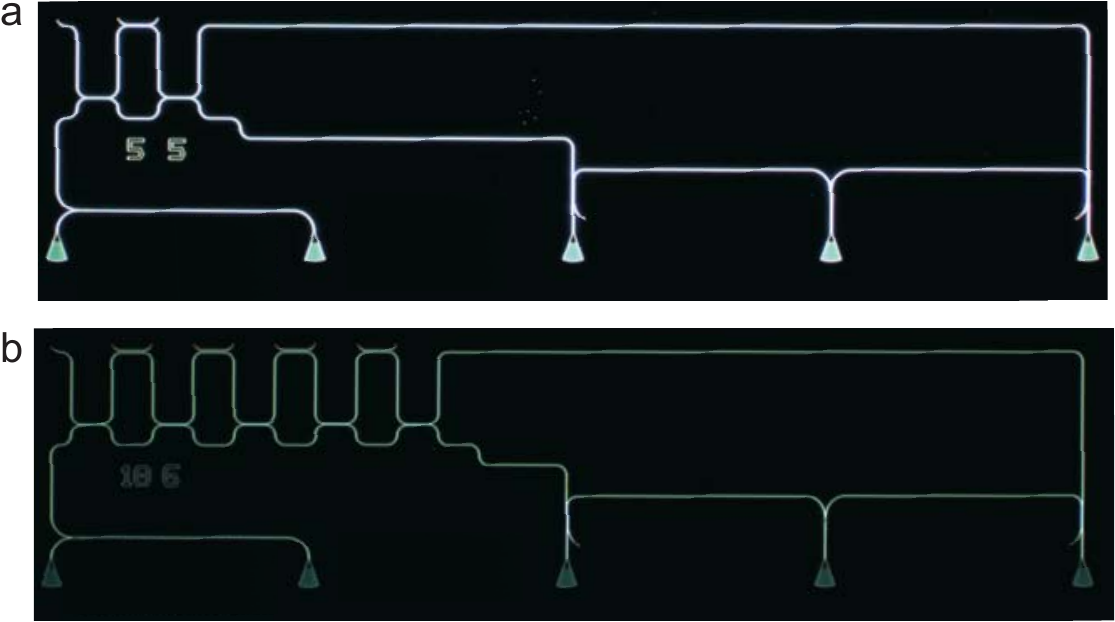}}
\caption{The optical micrographs of typical two-stage device (\textbf{a}) for IFM, and five-stage device (\textbf{b}) for QZIFM.} \label{photos}
\end{figure}

Here we show the optical micrographs of a typical two-stage MZI device for IFM in Fig.\ \ref{photos}\textbf{a}, and a five-stage device for QZIFM Fig.\ \ref{photos}\textbf{b}.

\section{The QZIFM device with 20 directional couplers}
\begin{figure}[h!]
\centerline{\includegraphics[width=0.5\textwidth]{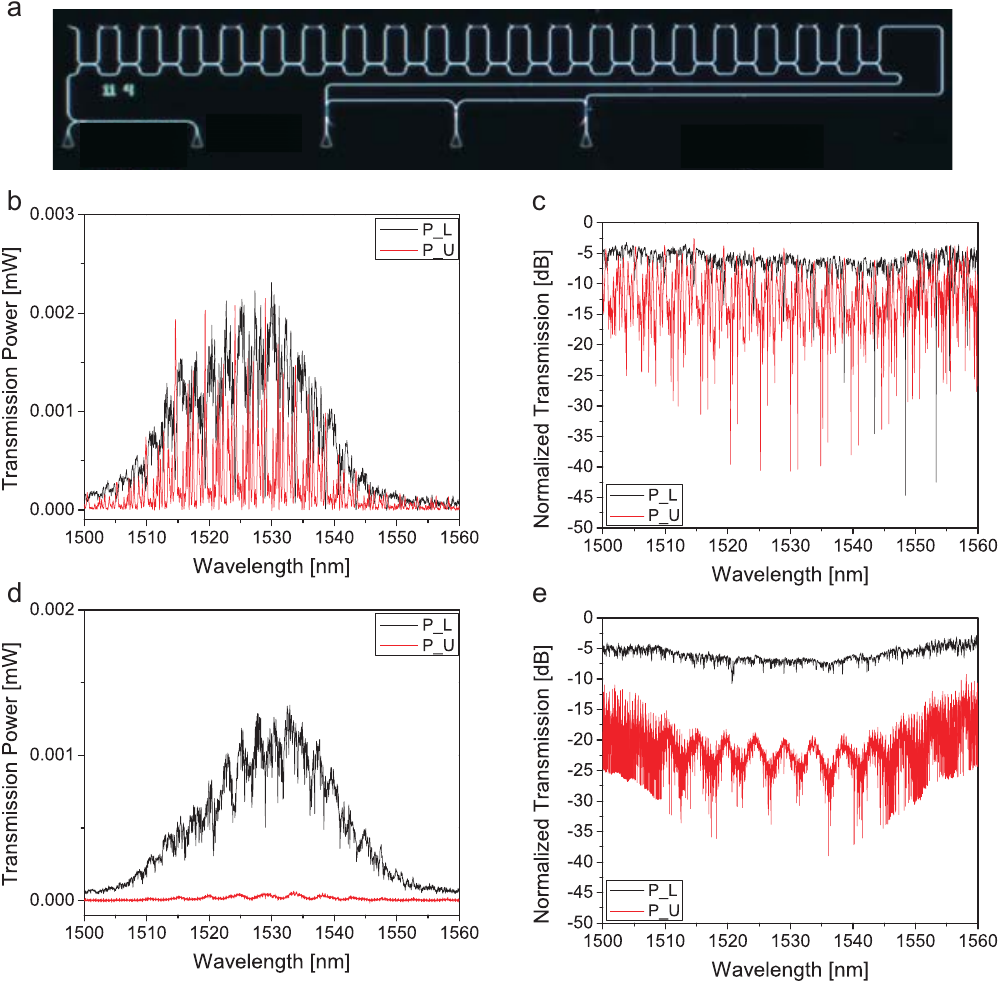}}
\caption{\textbf{a}. The optical micrograph of a 20-stage device for a QZIFM. \textbf{b}. The transmission spectra of the lower and upper outputs of a 20-stage device without absorber, which corresponds to the situation shown in Fig.\ 2\textbf{a} in the main text. \textbf{c}. The normalized transmission spectra in logarithm scale. \textbf{d}. The transmission spectra with 19 full absorbers being present in the upper arm of the interferometer. Note that this device has the same design as \textbf{b} except the gaps between absorbers and upper arm of connected interferometers have changed from $10 \mu$m to $190$ nm. \textbf{e}. The normalized transmission spectra in logarithm scale. It is clear to see that the interference disappeared, which is the signature of QZIFM.} \label{20stage}
\end{figure}

We have fabricated QZIFM devices with 20 directional couplers, in which we aim for realizing directional couplers with reflectivity $cos(\pi/40)^2=0.9938$ each. Based on Fig.\ \ref{design}\textbf{e}, the gap should be around $0.59 \mu m$ in this case. Experimentally, we found this gap to be $0.532 \mu m$ as the nominal value. In Fig.\ \ref{20stage}\textbf{a}, we show the optical micrograph of a 20-stage QZIFM device. In Fig.\ \ref{20stage}\textbf{b} and \textbf{c}, we show the transmission spectra of the 20-stage QZIFM device without absorbers in linear and logarithm scales, respectively. In Fig.\ \ref{20stage}\textbf{d} and \textbf{e}, we show the transmission spectra of the device with 19 full absorbers. in comparison with Fig.\ \ref{20stage}\textbf{b} and \textbf{c}, interference patterns disappear which is the signature of successful IFM. Note that the loss in the 20-stage QZIFM device is higher than that of 10-stage QZIFM device, which resulted lower IFM efficiency (shown in Fig.\ 5\textbf{e} of the main text).

\bibliographystyle{apsrev}

\end{document}